\documentclass[12pt]{article}
\usepackage{latexsym, amsbsy, amssymb,multirow, epsfig, amsmath}
\usepackage{hyperref}
\usepackage{epstopdf}
\usepackage{float}
\usepackage{tabularx}

\usepackage{natbib,graphicx,setspace,lscape,longtable,bm}
\usepackage{natbib,epsfig,graphicx}
\usepackage{amsmath,amsthm,amssymb,color}
\RequirePackage[mathlines, displaymath]{lineno}
\bibpunct{(}{)}{;}{a}{,}{,}

\newcommand{\csection}[1]
    {\begin{center}
        \stepcounter{section}
        {\bf\large\arabic{section}. #1}
    \end{center}
    \vspace{-0.15 cm}
}

\newcommand{\scsection}[1]
    {\begin{center}
        {\bf\large #1}
    \end{center}
    \vspace{-0.15 cm}
}

\newcommand{\csubsection}[1]{
\vspace{-0.25 cm}
\begin{center}
\stepcounter{subsection}
{\it\arabic{section}.\arabic{subsection}. #1}
\end{center}
\vspace{-0.25 cm}
}

\newcommand{\scsubsection}[1]{
\vspace{-0.25 cm}
\begin{center}
\stepcounter{subsection}
{\it #1}
\end{center}
\vspace{-0.25 cm}
}

\def\beqr{\begin{eqnarray}}
\def\eeqr{\end{eqnarray}}
\def\beqrs{\begin{eqnarray*}}
\def\eeqrs{\end{eqnarray*}}
\def\bep{\begin{prop}}
\def\eep{\end{prop}}
\def\be{\begin{equation}}
\def\ee{\end{equation}}
\def\bea{\begin{eqnarray}}
\def\eea{\end{eqnarray}}

\def\mR{\mathbb{R}}
\def\calN{\mathcal{N}}
\def\calA{\mathcal{A}}

\def\calP{\mathcal{P}}

\def\pr{\mbox{Pr}}

\def\var{\mbox{\rm var}}
\def\cov{\mbox{\rm cov}}
\def\corr{\mbox{\rm corr}}

\numberwithin{equation}{section}

\def\wh{\widehat}

\def\bb{\mbox{\boldmath$\beta$}}

\def\B{{\bf B}}

\def\v{{\bf v}}

\def\x{{\bf x}}

\def\C{{\bf C}}

\newcommand{\trans}{^{\mbox{\tiny{T}}}}
\def\defby{\stackrel{\mbox{\textrm{\rm\tiny def}}}{=}}

\newcommand{\bSig}{\mbox{\boldmath $\Sigma$}}

\newtheorem{theo}{\bf Theorem}
\newtheorem{lemma}{Lemma}
\newtheorem{prop}{\sc Proposition}

\begin{document}

\begin{center}
{\Large\bf Model-Free Conditional Feature Screening with Exposure Variables}\\
\bigskip
\bigskip
Yeqing Zhou, Jingyuan Liu*, Zhihui Hao and Liping Zhu
 \footnotetext{Yeqing Zhou is Ph.D. Student, School of Statistics and Management, Shanghai University of Finance and Economics, 777 Guoding Road,
		Shanghai 200433, P. R. China. Jingyuan Liu is Associate Professor, Department of Statistics in School of Economics, Wang Yanan Institute for Studies in Economics (WISE) and Fujian Key Laboratory of Statistical Science, Xiamen University, 422 South Siming Road, Xiamen 361005, P. R. China. Zhihui Hao is master student, WISE, Xiamen University, 422 South Siming Road, Xiamen 361005, P. R. China. Liping Zhu is Professor, Research Center for Applied Statistical Science and Institute of Statistics and Big Data, Renmin University of China,
		59 Zhongguancun Avenue, Haidian District,
		Beijing 100872, P. R. China.
		
		This work is supported by National Natural Science Foundation of P. R. China
		(11371236, 11771361 and 11422107),
		Fundamental Research Funds for the Scientific Research Foundation for the Returned Overseas Chinese Scholars, Ministry of Education, P. R. China  and
		Henry Fok Education Foundation Fund of Young College Teachers (141002).

* Corresponding to jingyuan@xmu.edu.cn.}\\
This version: \today
\end{center}
\begin{singlespace}
\begin{abstract}
In high dimensional analysis, effects of explanatory variables on responses sometimes rely on certain exposure variables, such as time or environmental factors. In this paper, to characterize the importance of each predictor, we utilize its conditional correlation given exposure variables with the empirical distribution function of response. A model-free conditional screening method is subsequently advocated based on this idea, aiming to identify significant predictors whose effects may vary with the  exposure variables. The proposed screening procedure is applicable to any model form, including that with heteroscedasticity where the variance component may also vary with exposure variables. It is also robust to extreme values or outlier. Under some mild conditions, we establish the desirable sure screening  and the ranking consistency properties of the screening method. The finite sample performances are illustrated by simulation studies and an application to the breast cancer dataset.

\noindent{\bf KEY WORDS:}
conditional screening ;
feature screening;
sure screening property;
variable selection.
\end{abstract}
\end{singlespace}

\newpage
\csection{Introduction}
 Ultrahigh dimensional data arise in many frontier areas, such as genetics, imaging, economics and finance. In these areas, quite often tremendous amount of explanatory variables are collected, while only a few  predictors are truly important to the response. To identify these truly active predictors, a variety of variable selection methods were studied based on different models.  One appealing method to select important variables and reduce the predictor dimensionality is the two-stage approach: feature screening methods are first conducted to roughly rule out the marginally unimportant predictors, and subsequent regularized regression approaches are applied to recover the final sparse models. In the screening stage, \cite{fan2008sure} first proposed a  sure independent screening procedure (SIS) based on the marginal Pearson correlation in the context of linear models. Its appealing sure screening property urges statisticians to extend the idea of SIS under different settings, including the generalized linear models \citep{fan2010sure}, semiparametric models \citep{li2012robust}, nonparametric additive models \citep{FanFengSong:2011, he2013quantile}. \cite{fan2010sure} stated that the validity of screening methods usually relies on the correct underlying-model specification, which motivates researchers to propose screening methods at a model-free basis, such as SIRS \citep{zhu2011model} and DC-SIS \citep{li2012feature}. See  \cite{Liu2015Aselective} for a selective overview of the screening methods.

 However, the effects of predictors on the response are sometimes dependent on certain exposure variables  in an unknown pattern, such as time or some environmental indices. For instance, in human genetics research, gene effects on certain phenotype, say body mass index, may reply on the current age of people. 
 When the predictors affect the response via one or more exposure variables, the corresponding effects  are often depicted by the interactions between predictors and the exposure variables  in linear models, or by the nonparametric coefficient functions in varying coefficient models.
 In the varying coefficient model, the dependence between predictors and response can be marginally assessed by the conditional Pearson correlation given exposure variables, due to the linearity of varying coefficient models given exposure variables. Therefore, to reduce the dimensionality of such ultrahigh dimensional varying coefficient models, \cite{Liu2014Feature}  and \cite{Fan2014Nonparametric} studied several conditional  screening methods, based on such the partial correlation and the conditional correlation learning.

 In analysis of ultrahigh dimensional data, unfortunately, we are often lack of prior information on the regression structure \citep{zhu2011model}, and the aforementioned linearity can be easily violated. In addition, extreme values or outliers often take a non-negligible role when tremendous amount of data are collected, destroying the nice data structure for applying the methods to the well-designed varying coefficient models. Under some other circumstances, predictors might be responsible for the variance, rather than the mean of response and the exposure variables may also play a role in the effects on the variance component. To address these issues, \cite{wen2018sure} developed a model-free screening method based on conditional distance correlation learning \citep{wang2015conditional}. However,
 the performance of this method is easily influenced by the presence of extreme values or outliers in the observations. Thus, we are motivated to use a robust metric to measure the conditional association between the predictors and response given exposure variables and apply it to the feature screening procedures. We adopt the conditional correlation between the predictor and  indicator function of response given the exposure variables. It employs the conditional rank instead of the original observed value of the response and thus stays invariant after strictly monotone transformation of the response. In estimation, the standard Nadaraya-Watson estimator is applied, which is easy to implement. Using the metric as a marginal score function, we further develop a model-free conditional sure independence screening procedure. The sure screening property \citep{fan2008sure} and ranking consistency property \citep{zhu2011model} of the screening procedure are carefully studied. We conduct extensive simulations to illustrate our
proposed method is effective to detect both linear and nonlinear conditional relations between the predictors and response given exposure variables, and ranks the important predictors above the unimportant ones with an overwhelming probability.

The rest of this paper is organized as follows. In Section 2, we propose a model-free conditional feature screening procedure based on the correlation learning, with a careful study of its theoretical properties. In Section 3,  we conduct Monte Carlo simulations to evaluate the finite sample performance of our proposals, and apply the method to analyze the breast cancer data. A discussion is given in Section 4. All technical proofs are relegated to the Appendix.

\csection{Conditional Sure Independence Ranking and Screening}
\csubsection{Some Preliminaries}
Suppose $Y\in\mR^1$ is the response variable, $\x = (X_1,\ldots, X_p)\trans\in\mR^p$ is the associated predictor vector and $u\in\mathbb{U}$ is the exposure variable. Given the exposure variables $u$, we define the set of active predictors  without model-specification:
\begin{equation}\label{true}\calA \defby \{k: \textrm{Given} \ u \in \mathbb{U}, \ F(y \mid \x,u) \ \textrm{varies with} \ X_k \ \textrm{for some} \ y \in\mR^1\},\end{equation}
where $F(y \mid \x,u)$ stands for the conditional distribution function of $Y$ given $\x$ and $u$. (\ref{true}) indicates that the truly active predictors affect the response variable through its distribution function, which may also depend on $u$. The set of inactive predictors is denoted by $\calA^c$, the complementary set of $\calA$. A screening method aims at removing as many predictors inactive predictors $\x_{\calA^c}$ as possible while retaining all the active predictors $\x_{\calA}$. Thus, we need to adopt a reasonable metric to measure the relative importance of each predictor conditioning on the exposure variables $u$.

We briefly review the sure independent ranking and screening procedure \citep[SIRS]{zhu2011model}, which identifies active predictors satisfying $F(y\mid \x)=F(y\mid \x_\calA)$ for all $ y \in \mR^1$. For easy illustration of its rationale, assume that $\x$ follows standard multivariate normal distribution and each predictor is standardized. The conditional distribution $Y$ given $\x$ varies with $\x_\calA$, and stay constant with $\x_\calA^c$. Thus it is natural to expect that $E\{\partial F(y \mid X_k)/ \partial X_k\}$ to be non-zero for $ k \in \calA$ and zero for $ k \in \calA^c$. The
normality assumption implies that $E\{\partial F(y \mid X_k)/ \partial X_k\}=E\{X_k I(Y \le y)\}$, where $I(\cdot)$ is an indicator function. Thus,  by defining $\rho_k(y)\defby E\{X_k I(Y \le y)\}$,  \cite{zhu2011model} employs $E\{\rho^2_k(Y)\}$ to rank the relative importance of predictors. The indicator function in $\rho_k(y)$ ensures the robustness of the method to extreme values and outliers.

When exposure variable $u$ is involved, however, the distribution of $Y$, as well as its association with the predictors, may vary with $u$. Under this circumstance, only considering the marginal expectations in $\rho_k(y)$ may miss important $u$-varying information. Instances indeed exist \citep{Liu2014Feature} where marginal screening procedures fail to detect those predictors with varying effects of $u$. To address this issue, we define the conditional correlation between the predictor $X_k$ for $k=1,\ldots,p$ and the indicator function of response $Y$ conditioning on $u$ as follows.
\beqrs
\Omega_k(u,y)\defby \corr(X_k,I(Y\leq y)\mid u)=\frac{\cov(X_k,I(Y\leq y)\mid u)}{\sqrt{\var(X_k\mid u)\var(I(Y\leq y)\mid u)}}.
\eeqrs
Then the marginal utility for screening becomes
\beqrs
\omega_k=E_y\left[E_u\left\{\Omega^2_k(u,y)\right\}\right],\ k=1,\ldots,p.
\eeqrs
To estimate $\omega_k$ based on the random sample $\{(\x_i,Y_i, u_i),i=1,\ldots,n\}$, we adopt Nadaraya-Watson estimator for each conditional mean used to compute $\Omega_k(u,y)$. Sepecially,
\beqrs
\wh E(X_k\mid u_j)&=&n^{-1}\sum_{i=1}^n K_h(u_i-u_j)X_{ik}/n^{-1}\sum_{i=1}^n K_h(u_i-u_j),\\
\wh E(X^2_k\mid u_j)&=&n^{-1}\sum_{i=1}^n K_h(u_i-u_j)X^2_{ik}/n^{-1}\sum_{i=1}^n K_h(u_i-u_j),\\
\wh E\{I(Y \le Y_l )\mid u_j\}&=&n^{-1}\sum_{i=1}^n K_h(u_i-u_j)I(Y_i \le Y_l)/n^{-1}\sum_{i=1}^n K_h(u_i-u_j),\\
\wh E\{X_kI(Y \le Y_l )\mid u_j\}&=&n^{-1}\sum_{i=1}^n K_h(u_i-u_j)X_{ik}I(Y_i \le Y_l)/n^{-1}\sum_{i=1}^n K_h(u_i-u_j),
\eeqrs
where $K_h(u) = K(u/h)/h$, $K(\cdot)$ is a kernel function and $h$ is the bandwidth. Then a natural estimator of $\hat{\omega}_k$ is
\beqrs
\hat{\omega}_k=\frac{1}{n^2}\sum_{l=1}^n\sum_{j=1}^n\frac{\widehat{\cov}^2\{X_k,I(Y\leq Y_l)\mid u_j\}}{\widehat{\var}(X_k\mid u_j)\widehat{\var}\{I(Y\leq Y_l)\mid u_j\}},
\eeqrs
where $\widehat{\cov}\{X_k,I(Y\leq Y_l)\mid u_j\}$ can be estimated through $\wh E\{X_kI(Y \le Y_l )\mid u_j\}-\wh E(X_k\mid u_j)\wh E\{I(Y \le Y_l )\mid u_j\}$. The variance term can be estimated by the similar patterns.
Based on the sample estimation of $\hat{\omega}_k$, we conduct the screening criterion to identify the active set indexed by
\beqr
\widehat{\calA}=\{k: \hat{\omega}_k \mbox{ ranks in the top $d$ among all $\hat{\omega}$'s}, \textrm{ for } 1\leq k\leq p\},
\eeqr
where $d$ is the user-specified threshold value. We refer the proposed conditional sure independent ranking and screening procedure as C-SIRS in the paper.

\csubsection{Theoretical Properties}
We establish several appealing properties for our proposed screening procedure. Denote $\lambda_{\max}(\C)$ and $\lambda_{\min}(\C)$ for the largest and smallest eigenvalues of a matrix $\C$, respectively. Write $\v^2=\v\v^{\trans}$ for a vector $\v$, and $\Omega_{\calA}(u,y)\defby\corr\{\x_{\calA},I(Y \le y )\mid u\}$. Say $a_n-b_n$ uniformly in $n$ if $\liminf\limits_{n\longrightarrow \infty}{a(n)-b(n)}>0$.

\noindent The following three conditions are required for Theorem 1.
\begin{enumerate}
	\item[(A1)] $E\{\x \mid \x_{\calA}^{\trans}\bb_{\calA}(u),u\}=\cov\{\x,\x_{\calA}^{\trans} \mid u\}\bb_{\calA}(u)[\cov\{\x_{\calA}^{\trans}\bb_{\calA}(u) \mid u\}]^{-1}\bb^{\trans}_{\calA}(u)\x_\calA$.
	\item[(A2)] $\min\limits_{k \in \calA} \omega_k>
	E\Bigg[ \lambda_{\max}\left\{
	\cov(\x_\calA,\x_{\calA^c}^{\trans} \mid u)\cov(\x_{\calA^c} ,\x_\calA^{\trans}\mid u) \right\} \lambda_{\max}\{\Omega_{\calA}^2(u,y)\}$\\
	$/\lambda^2_{\min}\left\{\cov(\x_\calA,\x_{\calA}^{\trans}\mid u)\right\}\Bigg]$ holds uniformly for $n$.
	\item[(A3)] $\x $ and $Y$ are independent conditioning on $\x^{\trans}_\calA \bb_\calA(u), u$.
\end{enumerate}
Condition (A1) is referred to as the conditional linearity condition. Condition (A2) is the crucial assumption to guarantee the satisfactory performance of our proposal, which requires the minimal signal of the active predictors not too small. It also does not allow strong correlation between $\x_\calA$ and $\x_{\calA^c}$, or among $\x_\calA$ themselves given the exposure variable $u$. Note that (A2) holds automatically if $\x_\calA$ and $\x_{\calA^c}$ are uncorrelated conditioning on $u$. Similar conditions are assumed in \cite{zhu2011model} and \cite{Liu2014Feature}. Condition (A3) dictates that $Y$ relies on $\x$ via the linear combinations $\x^{\trans}_\calA \bb_\calA(u)$.
{\theo\label{theo:1} Suppose conditions (A1), (A2) and (A3) hold, then we have
	\beqrs\label{rank}\liminf\limits_{n\longrightarrow \infty} \left\{\min\limits_{k \in \cal{A}}  \omega_k -\max\limits_{k \in \calA^c} \omega_k\right\} >0.
	\eeqrs}
Theorem \ref{theo:1}  illustrates that signals between the important predictors and the unimportant ones are distinguishable, which is a prerequisite for the ranking consistency property.

We  assume the following regularity conditions to derive the theoretical properties of C-SIRS.
Define $E(\cdot \mid u)=g(\cdot \mid u)/f(u)$.

\begin{enumerate}
	\item[(C1)]\textit{(The Kernel Function)}
	The kernel $K(\cdot)$ is a density function with compact support.  It is symmetric about zero and Lipschitz continuous. In addition,  it satisfies
	\beqrs
	\int_{-1}^1K(t)dt=1,\quad \int_{-1}^1t^{i-1}K(t)dt=0, \, 0 \le i \le m-1, \quad 0 \neq\int t^{m}K(t)dt=\nu_m<\infty
	\eeqrs
	It is bounded uniformly such that $\sup\limits_{u \in \mathbb{U}}|K(u)| \leq M_K < \infty$.
	\item[(C2)]\textit{(The Density)} The probability density functions of $u$, denoted by $f(u)$ has continuous second-order derivative on $\mathbb{U}$.
	\item[(C3)]\textit{(The Derivatives)} The $(m-1)$-th derivatives of  both $g(\cdot \mid u)$, $f(u)$ are locally Lipschitz-continuous with respect to $u$.
	\item[(C4)]\textit{(The Bandwidth)} The bandwidth  $h$  satisfies $h=O(n^{-\theta})$, for some $\theta$   which satisfies $(4m)^{-1} < \theta < 1/4$.
	\item[(C5)]\textit{(The Moment Condition)} There exists a positive constant $s_0$ such that
	\beqrs
	\sup\limits_{u \in \mathbb{U}}\max_{1 \le k \le p} E\{\exp(sX^2_k)\mid u\} < \infty, \ \textrm{for} \ 0<s \le s_0.
	\eeqrs
	Further assume that $E(X_k|u)$ and $E(X^2_k|u)$, their first-order and second-order derivatives are finite uniformly in $u \in \mathbb{U}$.
\end{enumerate}
{\theo\label{theo:2} (Sure Screening Property) Under the  conditions (C1)-(C5), for any $0< \gamma + \theta \le 1/4$ and $0< \gamma\le m\theta $, if $p$ satisfies
	$ n^3p\exp(  - c n^{1/2-2\gamma-2\theta}) \to 0$ and  $\min\limits_{k\in \mathcal{A}} \ \omega_k \geq 2cn^{-\gamma}$ for some  $c>0$, then
	\beqrs\label{sure}
	\pr\big(\mathcal{A} \subseteq \widehat{\mathcal{A}}\big) \geq 1- O\left\{n^{3}|\mathcal{A}| \exp(  - c n^{1/2-2\gamma-2\theta } )\right\},
	\eeqrs
	where $c$ is a generic constant and $|\mathcal{A}|$  is the cardinality of $\calA$.}

The sure screening property \citep{fan2008sure} of the C-SIRS procedure ensures that all truly active predictors can be retained after screening with the probability approaching to one.

{\theo\label{theo:3} Under the  conditions (C1)-(C5), in addition to conditions (A1)-(A3), if $p$ satisfies $ {n^3p\exp(  - c n^{1/2-2\theta})} \to 0$ for some  $c>0$, then  we have
	\beqrs
	\liminf\limits_{n\longrightarrow \infty} \left\{\min\limits_{k \in \cal{A}} \wh\omega_k -\max\limits_{k \in \calA^c} \wh\omega_k\right\} >0 \:\text{in probability}.
	\eeqrs}
The ranking consistency guarantees the active predictors ranked in the top, prior to the inactive ones, with an overwhelming probability.

\csection{Numerical Studies}
\csubsection{The Performance of Conditional Screening}

In this section, we investigate the finite sample performance of our proposed screening procedure through Monte Carlo simulations, and also compare it with three screening methods including SIRS \cite{zhu2011model}, DC-SIS \citep{li2012feature} and CC-SIS \citep{Liu2014Feature} and CDC-SIS \citep{wen2018sure}. Under all model settings, we  draw $\x=(X_1,X_2,\cdots, X_{p})\trans$ and an intermediate variable $u^*$ from the multivariate normal distribution with mean zero and AR covariance matrix $\bSig=(0.5^{|i-j|})_{(p+1)\times (p+1)}$. Then the exposure variable $u$ is obtained from $u\defby\Phi(u^*)$, where $\Phi(\cdot)$ is cumulative distribution function of the standard normal distribution $\calN(0,1)$.  We set the sample size $n=200$ and fix $p=1000$. Each experiment is repeated 1000 times. We adopt the Epanechnikov kernel $K(u)=0.75(1-u)_{+}$ in both simulations and real data analysis.

We evaluate the finite-sample performance through the following four criteria:
\begin{enumerate}
	\item $R_k$: The average of the ranks of  each important predictor out of 1000 replications.
	\item $\mathcal{S}$: The minimum model size  to ensure that all important predictors are included after screening. We expect it to be as close as the the number of truly active predictors. We report the 5$\%$, 25$\%$, 50$\%$, 75$\%$ and 95$\%$ quantiles of $\mathcal{S}$ out of 1000 replications.
	\item $\calP_a$: The proportion of all active predictors selected after screening for a given model size out of 1000 replications. We consider three screened model size $d=\nu[n^{4/5}/\log(n^{4/5})]$ varying $\nu=1,2$ and 3. The corresponding $d$ is 16, 32 and 48, respectively. We expect it to be as close to one as possible.
	\item $\calP_k$: The proportion of each active predictor selected after screening for a given model size out of 1000 replications.
\end{enumerate}
\noindent  {\bf Example 1.}  We first  generate the response $Y$ from the generalized varying coefficient models respectively:
 \begin{itemize}
	\item \emph{Case 1}: $\textrm{logit}\left\{E(Y\mid \mathbf{x},u)\right\}=\x^T\bb(u)$;
	\item \emph{Case 2}: $\log \left\{E(Y\mid \mathbf{x},u)\right\}=\x^T\bb(u)$,
\end{itemize}
where $\bb=\{\beta_1(u),\ldots,\beta_{1000}(u)\}^T$. $\beta_j$ is nonzero when $X_j$ is the active predictor and remains zero otherwise. We set he active predictors index to be $\{2,100,400,600,1000\}$, with corresponding coefficient functions $\beta_2(u)=2I(u>0.4), \:\beta_{100}(u)=1+u, \:\beta_{400}(u)=(2-3u)^2,\beta_{600}(u)=2\sin(2\pi u), \:\beta_{1000}(u)=\exp\{u/(u+1)\}$. The first model is  logistic varying coefficient model while the second one is the Poisson varying coefficient model.

\begin{table}[h]\footnotesize
	\begin{center}
		\caption{The mean of $R_k$ of each true predictor for Example 1.}\label{t1R}
		\setlength{\tabcolsep}{4mm}{
			\begin{tabular}{ccrrrrr}
				\hline
				&Method & $R_2$ & $R_{100}$ & $R_{400}$ & $R_{600}$ & $R_{1000}$ \\
				\hline
				\multirow{5}{*}{Case1}&SIRS & 30.89 & 8.22 & 109.32 & 491.70 & 18.37 \\
				& DC-SIS & 18.11 & 6.09 & 67.89 & 512.58 & 10.73 \\
				&CDC-SIS&18.25 & 4.34 & 61.82 & 457.82 & 10.26\\
				&CC-SIS & 9.85 & 9.63 & 7.63 & 2.49 & 17.68\\
				&C-SIRS & 10.41 & 7.15 & 5.69 & 3.02 & 14.34 \\
				\hline
				\multirow{5}{*}{Case2}	& SIRS & 7.05 & 2.44 & 123.44 & 462.06 & 3.44 \\
				&  DC-SIS & 17.53 & 3.66 & 163.17 & 164.99 & 6.36 \\
				&CDC-SIS&104.78 & 76.95 & 203.15 & 142.34 & 107.58 \\
				& CC-SIS & 21.81 & 10.00 & 65.09 & 1.56 & 19.07 \\
			    & C-SIRS & 3.96 & 4.67 & 35.12 & 1.08 & 6.75 \\
				\hline
		\end{tabular}}
	\end{center}
\end{table}

\begin{table}[h]\footnotesize
	\begin{center}
		\caption{ The quantiles of the minimum model size $\cal{S}$ for Example 1.}\label{t1M}
		\setlength{\tabcolsep}{4mm}{
			\begin{tabular}{ccrrrrr}
				\hline
				&Method & 5\% & 25\% & 50\% & 75\% & 95\% \\
				\hline
				\multirow{5}{*}{Case1}& SIRS & 77.00 & 277.75 & 513.00 & 753.00 & 953.00 \\
				& DC-SIS & 83.00 & 285.75 & 526.00 & 773.00 & 959.00 \\
				&CDC-SIS&75.95 & 245.75 & 463.00 & 691.25 & 898.00 \\
				&CC-SIS & 5.00 & 7.00 & 12.00 & 27.00 & 137.05 \\
				
				& C-SIRS & 5.00 & 6.00 & 10.50 & 22.00 & 96.05 \\
				\hline
				\multirow{5}{*}{Case2}	&SIRS &  75.90 & 255.75 & 488.00 & 725.75 & 942.05 \\
				&DC-SIS & 33.00 & 93.75 & 206.00 & 389.25 & 798.05 \\
			&CDC-SIS&69.00 & 191.75 & 336.50 & 547.25 & 810.20\\
				&CC-SIS & 7.00 & 19.00 & 45.00 & 116.00 & 345.10 \\
				&C-SIRS & 5.00 & 7.00 & 12.00 & 32.25 & 169.15 \\
				\hline
		\end{tabular}}
	\end{center}
\end{table}

\begin{table}[h]\footnotesize
	\begin{center}
		\caption{The proportions of $\calP_a$ and $\calP_k$ given the model size $d$ for Example 1.}\label{t1p}
		\setlength{\tabcolsep}{4mm}{
			\begin{tabular}{ccccccccc}
				\hline
				&$d$& Method& $\calP_2$ & $\calP_{100}$ & $\calP_{400}$ & $\calP_{600}$ & $\calP_{1000}$ & $\calP_a$ \\
				\hline
				\multirow{15}{*}{Case1}&\multirow{5}{*}{16}& SIRS & 0.71 & 0.91 & 0.38 & 0.01 & 0.81 & 0.00 \\
				& &DC-SIS &  0.81 & 0.93 & 0.52 & 0.01 & 0.87 & 0.00 \\
				&&CDC-SIS&0.82 & 0.96 & 0.56 & 0.01 & 0.90 & 0.00\\
				&&CC-SIS & 0.90 & 0.91 & 0.91 & 0.99 & 0.83 & 0.60 \\
				& & C-SIRS & 0.90 & 0.93 & 0.95 & 0.99 & 0.84 & 0.65 \\
				
				&\multirow{5}{*}{32}& SIRS  & 0.79 & 0.95 & 0.52 & 0.03 & 0.88 & 0.01 \\
				&  &DC-SIS & 0.90 & 0.97 & 0.64 & 0.03 & 0.93 & 0.01 \\
				&&CDC-SIS&0.88 & 0.98 & 0.67 & 0.02 & 0.94 & 0.01\\
				& &CC-SIS & 0.95 & 0.95 & 0.96 & 1.00 & 0.89 & 0.77 \\
				& & C-SIRS & 0.96 & 0.96 & 0.98 & 1.00 & 0.91 & 0.81 \\
				
				&\multirow{5}{*}{48}&SIRS & 0.85 & 0.97 & 0.59 & 0.04 & 0.91 & 0.02 \\
				&  &DC-SIS & 0.91 & 0.98 & 0.72 & 0.05 & 0.95 & 0.03 \\
				&&CDC-SIS&0.91 & 0.99 & 0.75 & 0.04 & 0.96 & 0.02\\
				&&CC-SIS & 0.97 & 0.97 & 0.98 & 1.00 & 0.92 & 0.84 \\
				& & C-SIRS & 0.97 & 0.97 & 0.99 & 1.00 & 0.94 & 0.87 \\
				\hline
				\multirow{15}{*}{Case2} &\multirow{5}{*}{16}&SIRS & 0.93 & 0.99 & 0.33 & 0.02 & 0.98 & 0.00 \\
				&& DC-SIS & 0.80 & 0.98 & 0.28 & 0.15 & 0.94 & 0.02 \\
				&&CDC-SIS& 0.41 & 0.52 & 0.23 & 0.29 & 0.42 & 0.00\\
				&&CC-SIS & 0.75 & 0.87 & 0.45 & 0.99 & 0.79 & 0.19 \\
				& & C-SIRS & 0.98 & 0.98 & 0.66 & 1.00 & 0.94 & 0.59 \\
				&\multirow{5}{*}{32}&SIRS & 0.97 & 0.99 & 0.47 & 0.04 & 0.99 & 0.02 \\
				&&DC-SIS & 0.88 & 0.99 & 0.40 & 0.24 & 0.97 & 0.05 \\
				&&CDC-SIS&0.52 & 0.61 & 0.31 & 0.41 & 0.51 & 0.01\\
				&&CC-SIS & 0.86 & 0.93 & 0.60 & 1.00 & 0.88 & 0.41 \\
				&&C-SIRS & 0.99 & 0.99 & 0.78 & 1.00 & 0.97 & 0.74 \\
				&\multirow{5}{*}{48} &SIRS & 0.98 & 1.00 & 0.56 & 0.06 & 0.99 & 0.03 \\
				& &DC-SIS & 0.92 & 0.99 & 0.47 & 0.33 & 0.98 & 0.11 \\
			   &&CDC-SIS&0.59 & 0.67 & 0.36 & 0.47 & 0.57 & 0.02\\
				&&CC-SIS & 0.91 & 0.95 & 0.68 & 1.00 & 0.92 & 0.52 \\
				& &C-SIRS & 1.00 & 1.00 & 0.83 & 1.00 & 0.99 & 0.82 \\
				\hline
		\end{tabular}}
	\end{center}
\end{table}

We report the simulation results of $R_k$  and $\cal{S}$ in Table \ref{t1R} and Table \ref{t1M}, respectively. The proposed C-SIRS method outperforms other competitors under both model settings. The rank of each predictor is on the top while the median of $\cal{S}$ is  close to the  number of truly active predictors, indicating that our proposal achieves a high accuracy in ranking. CC-SIS also performs satisfactorily as it is designed for the varying  coefficient model.
Both SIRS and DC-SIS are not able to detect the predictor $X_{600}$. This is mainly because the expectation of $\beta_{600}(u)$ is zero, making the predictor $X_{600}$ but marginally independent but conditional related to the response. The simulation results of  selection proportions $\calP_a$ and $\calP_k$  are summarized in Table \ref{t1p}. The C-SIRS method selects all important predictors with high probability, indicating  its sure screening property. The $\calP_a$ of SIRS, DC-SIS and CDC-SIS are negligible even for the largest submodel size $d = 48$.

\noindent  {\bf Example 2.}  We then consider following  four models, in which the response $Y$ depends on the predictors  nonlinearly with a given $u$:
\begin{itemize}
	\item \emph{Case 1}: $Y=\exp\{\x^T\bb(u)\}+\varepsilon$;
	\item \emph{Case 2}: $Y=\exp\{\x^T\bb(u)+\varepsilon\}$;
	\item \emph{Case 3}: $Y=\beta_2(u)\exp(X_2)+\beta_{100}(u)X_{100}^3+2\beta_{400}(u)X_{400}I(X_{400}<2)+\beta_{600}(u)X_{600}
	+1.5\beta_{1000}(u)X_{1000}+\varepsilon$;
	\item \emph{Case 4}:
	$Y=\beta_2(u)X_2+\beta_{100}(u)X_{100}+\beta_{400}(u)X_{400}+\beta_{1000}(u)X_{1000}+2\exp\{\beta_{600}(u)X_{600}\}\varepsilon$,
\end{itemize}
where the $\bb(u)$ setting remains identical as  Example 1 and the error term is independently generated from $t(1)$ for Case 1 to Case 3 and $\calN(0,1)$ for Case 4. Notice that Case 4 demonstrates the heteroscedasticity issue, where the variance component is affected by $X_{600}$, and the effect vary with $u$.

\begin{table}[h]\footnotesize
	\begin{center}
		\caption{The mean of $R_k$ of each true predictor for Example 2.}\label{t2R}
		\setlength{\tabcolsep}{4mm}{
			\begin{tabular}{ccrrrrr}
				\hline
				& Method& $R_2$ & $R_{100}$ & $R_{400}$ & $R_{600}$ & $R_{1000}$ \\
				\hline
				\multirow{5}{*}{Case1}	&SIRS & 24.64 & 2.76 & 61.90 & 485.22 & 9.39 \\
				& DC-SIS & 104.67 & 65.33 & 277.81 & 171.04 & 146.13 \\
			    &CDC-SIS&108.22 & 79.91 & 206.46 & 143.12 & 111.44\\
				& CC-SIS & 188.26 & 167.32 & 131.75 & 56.50 & 169.04 \\
				&C-SIRS & 5.69 & 8.36 & 10.61 & 1.81 & 6.97 \\
				\hline
				\multirow{5}{*}{Case2}&SIRS & 22.78 & 3.29 & 53.03 & 478.24 & 7.70 \\
				&  DC-SIS & 362.10 & 520.40 & 560.70 & 579.80 & 538.20 \\
				&CDC-SIS& 497.71 & 484.76 & 506.99 & 500.14 & 487.19\\
				& CC-SIS & 231.46 & 197.64 & 143.85 & 71.32 & 188.68 \\
				&C-SIRS & 9.63 & 6.71 & 18.70 & 2.39 & 12.31 \\
				\hline
				\multirow{5}{*}{Case3}&SIRS & 9.39 & 1.24 & 31.41 & 439.37 & 7.57 \\
				
				&DC-SIS & 20.31 & 8.49 & 52.85 & 407.09 & 31.37 \\
				&CDC-SIS&18.45 & 7.61 & 50.79 & 389.78 & 29.05\\
				&CC-SIS & 123.35 & 34.00 & 151.27 & 197.23 & 177.44 \\
				&C-SIRS &  6.49 & 1.34 & 8.67 & 19.35 & 8.22 \\
				\hline
				\multirow{5}{*}{Case4}&	SIRS & 21.89 & 4.01 & 46.42 & 457.25 & 6.79 \\
				&DC-SIS & 226.48 & 172.32 & 295.15 & 5.51 & 202.70 \\
				&CDC-SIS& 234.92 & 175.71 & 318.88 & 4.99 & 204.87\\
				&CC-SIS & 452.48 & 465.33 & 486.95 & 74.69 & 469.96 \\
				&C-SIRS &  7.99 & 7.39 & 17.09 & 8.78 & 14.56 \\
				\hline
		\end{tabular}}
	\end{center}
\end{table}
\begin{table}[h]\footnotesize
	\begin{center}
		\caption{ The quantiles of the minimum model size $\cal{S}$ for Example 2.}\label{t2M}
		\setlength{\tabcolsep}{4mm}{
			\begin{tabular}{ccrrrrr}
				\hline
				&Method & 5\% & 25\% & 50\% & 75\% & 95\% \\
				\hline
				\multirow{5}{*}{Case1}&SIRS & 68.95 & 259.75 & 473.50 & 728.00 & 933.10 \\
				&DC-SIS & 82.00 & 196.75 & 359.00 & 583.25 & 844.05 \\
				&CDC-SIS&72.95 & 194.00 & 353.00 & 550.00 & 822.05\\
				&CC-SIS & 65.95 & 185.00 & 328.00 & 568.75 & 865.00 \\
				&C-SIRS & 5.00 & 6.00 & 9.00 & 21.00 & 92.00 \\
				\hline
				\multirow{5}{*}{Case2}&SIRS & 95.00 & 203.15 & 495.20 & 702.00 & 925.45 \\
				&DC-SIS & 42.75 & 126.25 & 251.50 & 505.20 & 879.55 \\
				&CDC-SIS&563.95 & 762.75 & 872.00 & 946.00 & 989.00 \\
				&CC-SIS & 45.00 & 162.00 & 346.00 & 578.95 & 882.00 \\
				&C-SIRS & 5.00 & 7.50 & 13.25& 42.50 & 199.65 \\
				\hline
				\multirow{5}{*}{Case3}&	SIRS & 43.90 & 194.00 & 392.00 & 667.00 & 920.10 \\
				
				&DC-SIS& 50.00 & 178.75 & 369.50 & 633.75 & 911.05 \\
				&CDC-SIS&42.00 & 174.00 & 376.00 & 589.50 & 892.05 \\
				&CC-SIS & 19.00 & 130.00 & 353.50 & 645.00 & 918.25 \\
				&C-SIRS & 5.00 & 6.00 & 10.00 & 23.00 & 114.00 \\
				\hline
				\multirow{5}{*}{Case4}&		SIRS & 59.95 & 237.00 & 436.50 & 689.25 & 933.00 \\
				
				&DC-SIS & 68.95 & 228.75 & 407.00 & 617.00 & 842.10 \\
				&CDC-SIS&98.00 & 267.75 & 442.50 & 629.00 & 844.05\\
				&CC-SIS & 432.90 & 673.75 & 811.50 & 918.00 & 984.00 \\
				&C-SIRS &6.00 & 9.00 & 16.00 & 36.00 & 154.00 \\
				\hline

		\end{tabular}}
	\end{center}
\end{table}
\begin{table}[ht!]\footnotesize
	\begin{center}
		\caption{The proportions of $\calP_a$ and $\calP_k$ given the model size $d$ for Example 2.}\label{t2p}
		\setlength{\tabcolsep}{4mm}{
			\begin{tabular}{ccccccccc}
				\hline
				&$d$& Method& $\calP_2$ & $\calP_{100}$ & $\calP_{400}$ & $\calP_{600}$ & $\calP_{1000}$ & $\calP_a$ \\
				\hline
				\multirow{15}{*}{Case1}&\multirow{5}{*}{16}&SIRS & 0.87 & 0.97 & 0.63 & 0.01 & 0.93 & 0.00 \\
				&&DC-SIS & 0.40 & 0.54 & 0.23 & 0.24 & 0.34 & 0.00 \\
				&&CDC-SIS&0.40 & 0.51 & 0.23 & 0.29 & 0.41 & 0.00\\
				&&CC-SIS & 0.18 & 0.28 & 0.32 & 0.63 & 0.22 & 0.00 \\
				&&C-SIRS & 0.92 & 0.97 & 0.90 & 1.00 & 0.91 & 0.72 \\
				&\multirow{5}{*}{32}&SIRS & 0.92 & 0.99 & 0.76 & 0.03 & 0.96 & 0.02 \\
				&&DC-SIS & 0.50 & 0.64 & 0.30 & 0.33 & 0.45 & 0.00 \\
				&&CDC-SIS&0.52 & 0.60 & 0.30 & 0.41 & 0.50 & 0.01\\
				&&CC-SIS & 0.28 & 0.38 & 0.48 & 0.75 & 0.33 & 0.01 \\
				&&C-SIRS & 0.98 & 0.99 & 0.97 & 1.00 & 0.96 & 0.90 \\
				&\multirow{5}{*}{48}&SIRS & 0.93 & 0.99 & 0.83 & 0.05 & 0.99 & 0.03 \\
				
				&&DC-SIS & 0.61 & 0.68 & 0.33 & 0.38 & 0.51 & 0.01 \\
				&&CDC-SIS&0.58 & 0.66 & 0.35 & 0.46 & 0.57 & 0.02\\
				&&CC-SIS & 0.37 & 0.44 & 0.53 & 0.78 & 0.44 & 0.01 \\
				&&C-SIRS & 1.00 & 0.99 & 0.99 & 1.00 & 0.98 & 0.96 \\
				\hline
				\multirow{15}{*}{Case2} &\multirow{5}{*}{16}&  SIRS & 0.82 & 0.97 & 0.61 & 0.02 & 0.92 & 0.01 \\
			
				&&  DC-SIS & 0.38 & 0.53 & 0.26 & 0.28 & 0.31 & 0.00 \\
				&&CDC-SIS&0.01 & 0.02 & 0.01 & 0.03 & 0.02 & 0.00\\
				&&CC-SIS &  0.16 & 0.23 & 0.34 & 0.59 & 0.21 & 0.01 \\
				&&C-SIRS & 0.92 & 0.93 & 0.81 & 0.99 & 0.86 & 0.58 \\
				&\multirow{5}{*}{32}&SIRS & 0.89 & 0.99 & 0.72 & 0.05 & 0.96 & 0.03 \\
			
				&&DC-SIS & 0.51 & 0.65 & 0.32 & 0.36 & 0.43 & 0.01 \\
				&&CDC-SIS& 0.03 & 0.04 & 0.03 & 0.05 & 0.04 & 0.00 \\
				&& CC-SIS & 0.23 & 0.36 & 0.45 & 0.71 & 0.34 & 0.01 \\
				&&C-SIRS & 0.95 & 0.97 & 0.89 & 1.00 & 0.92 & 0.74 \\
				&\multirow{5}{*}{48}&SIRS & 0.92 & 0.99 & 0.79 & 0.07 & 0.97 & 0.06 \\
				
				&&DC-SIS & 0.64 & 0.69 & 0.35 & 0.37 & 0.56 & 0.01 \\
				&&CDC-SIS&0.05 & 0.06 & 0.04 & 0.07 & 0.05 & 0.00 \\
				&&CC-SIS & 0.39 & 0.43 & 0.56 & 0.81 & 0.48 & 0.01 \\
				&&C-SIRS & 0.97 & 0.98 & 0.92 & 1.00 & 0.95 & 0.83 \\
				
				\hline
			
		\end{tabular}}
	\end{center}
\end{table}

\begin{table}[ht!]\footnotesize
	\begin{center}
		\caption{The proportions of $\calP_a$ and $\calP_k$ given the model size $d$ for Example 2.}\label{t2p-1}
		\setlength{\tabcolsep}{4mm}{
			\begin{tabular}{ccccccccc}
				\hline
				&$d$& Method& $\calP_2$ & $\calP_{100}$ & $\calP_{400}$ & $\calP_{600}$ & $\calP_{1000}$ & $\calP_a$ \\
				\hline
				\multirow{15}{*}{Case3} &\multirow{5}{*}{16}	&SIRS & 0.89 & 1.00 & 0.72 & 0.01 & 0.93 & 0.01 \\
				&& DC-SIS &  0.84 & 0.95 & 0.65 & 0.01 & 0.76 & 0.01 \\
				&&CDC-SIS&0.85 & 0.96 & 0.65 & 0.01 & 0.77 & 0.01\\
				&&CC-SIS &  0.50 & 0.84 & 0.42 & 0.23 & 0.28 & 0.03 \\
				&&C-SIRS & 0.93 & 1.00 & 0.92 & 0.78 & 0.93 & 0.63 \\
				&\multirow{5}{*}{32}&SIRS &  0.94 & 1.00 & 0.81 & 0.04 & 0.96 & 0.03 \\
				
				&&DC-SIS & 0.90 & 0.96 & 0.75 & 0.04 & 0.85 & 0.03 \\
				&&CDC-SIS& 0.90 & 0.97 & 0.73 & 0.05 & 0.85 & 0.03 \\
				&& CC-SIS & 0.58 & 0.86 & 0.51 & 0.32 & 0.40 & 0.09 \\
				&&C-SIRS & 0.97 & 1.00 & 0.96 & 0.88 & 0.97 & 0.79 \\
				&\multirow{5}{*}{48}	&SIRS & 0.96 & 1.00 & 0.86 & 0.07 & 0.97 & 0.06 \\
				
				&&DC-SIS & 0.92 & 0.97 & 0.79 & 0.06 & 0.88 & 0.05 \\
				&&CDC-SIS&0.93 & 0.98 & 0.79 & 0.08 & 0.87 & 0.06 \\
				&&CC-SIS & 0.62 & 0.88 & 0.55 & 0.38 & 0.46 & 0.12 \\
				&&C-SIRS &  0.98 & 1.00 & 0.97 & 0.91 & 0.98 & 0.85 \\
				
				\hline
				\multirow{15}{*}{Case4} &	\multirow{5}{*}{16}	& SIRS & 0.78 & 0.97 & 0.58 & 0.02 & 0.93 & 0.01 \\
				
				&&DC-SIS & 0.13 & 0.22 & 0.08 & 0.92 & 0.15 & 0.00 \\
				&&CDC-SIS&0.12 & 0.20 & 0.05 & 0.95 & 0.16 & 0.00\\
				&&CC-SIS & 0.03 & 0.02 & 0.02 & 0.59 & 0.02 & 0.00 \\
				&&C-SIRS & 0.91 & 0.95 & 0.80 & 0.87 & 0.87 & 0.50 \\
				&\multirow{5}{*}{32}&SIRS & 0.87 & 0.98 & 0.72 & 0.04 & 0.96 & 0.02 \\
				&&DC-SIS & 0.20 & 0.32 & 0.13 & 0.97 & 0.24 & 0.01 \\
				&&CDC-SIS&0.20 & 0.29 & 0.10 & 0.98 & 0.23 & 0.01\\
				&& CC-SIS & 0.05 & 0.05 & 0.04 & 0.68 & 0.05 & 0.00 \\
				&&C-SIRS & 0.95 & 0.97 & 0.88 & 0.96 & 0.92 & 0.72 \\
				&\multirow{5}{*}{48}&SIRS & 0.90 & 0.99 & 0.78 & 0.06 & 0.97 & 0.04 \\
				
				&&DC-SIS & 0.26 & 0.40 & 0.16 & 0.98 & 0.31 & 0.03 \\
				&&CDC-SIS&0.26 & 0.35 & 0.14 & 0.99 & 0.28 & 0.01 \\
				&&CC-SIS & 0.07 & 0.07 & 0.06 & 0.73 & 0.06 & 0.00 \\
				&&C-SIRS & 0.97 & 0.97 & 0.91 & 0.99 & 0.94 & 0.81 \\
				\hline
		\end{tabular}}
	\end{center}
\end{table}

The simulation results are tabulated in Table \ref{t2R} for $R_k$, Table \ref{t2M} for $\cal{S}$ and Table \ref{t2p}, \ref{t2p-1} for  $\calP_a$ and $\calP_k$.  It can be clearly seen that C-SIRS method is still the obvious winner, and performs well in both ranking and selection. CC-SIS method  is not able to capture the nonlinear conditional dependence, and the rankings are not accurate. The performances of DC-SIS and CDC-SIS methods are still poor, because of the heavy distribution of the response. We observe that the SIRS method also fails to identify the $X_{600}$, which is in accordance with Example 1.

\csubsection{Real data analysis}
Breast cancer is the one of most common  malignancy among women, with a high lethality rate. There is a urgent need for the early diagnosis of breast cancer and simultaneously monitoring the disease progression.  In this section, we evaluate the performance of proposed C-SIRS procedure through the breast cancer data reported by \cite{chin2006genomic}.   The dataset, available from the R package \cite{PMA:2013}, consists of 19672 gene expressions and 2149 CGH measurements over 88 cancer patient samples. Our interest is to detect the most influential genes to the comparative genomic hybridization (CGH) measurement, as in some previous research \citep{witten2009penalized,wen2018sure}. CGH is known as an indicator of the genome copy number variation and cancer diagnosis. In addition, recent researches \citep{mcpherson2000abc,desantis2017breast} found that age is a risk factor that influences the incidence of breast cancer. Thus, we treat age as the exposure variable $u$ in our analysis. Furthermore, we obtain the first principal component of 136 CGH measurements as the response, and the 19672  gene expressions as predictors.

\begin{figure}[!h]
	\begin{center}
		\includegraphics[width=15cm,height=12cm]{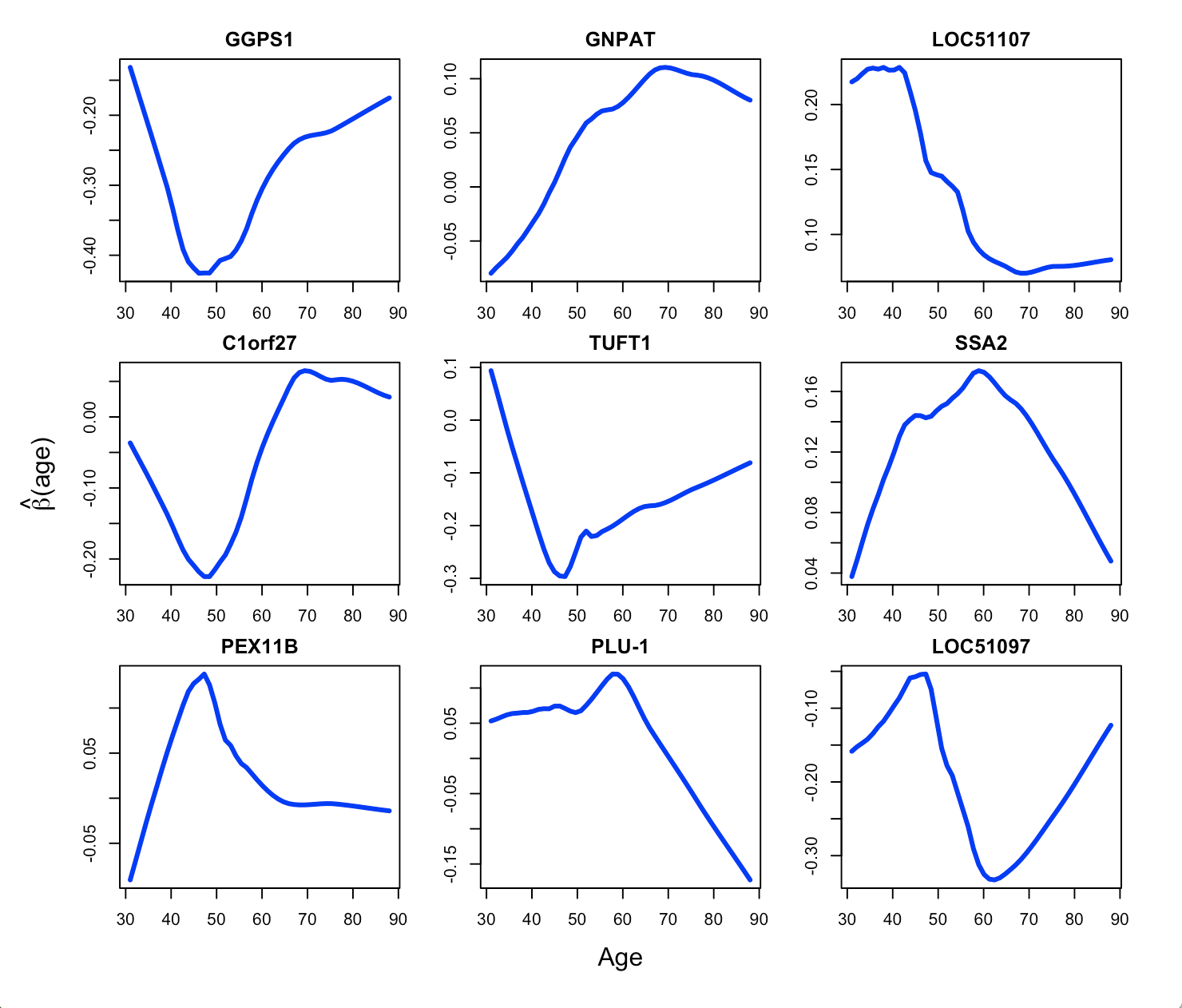}
	\end{center}
\caption{The estimated coefficient functions of the selected nine genes.}\label{coef}
\end{figure}

In our analysis, we first use the five screening methods SIRS, DC-SIS, CDC-SIS, CC-SIS and the proposed C-SIRS to select the most of relevant genes with top $d=2[n^{4/5}/\log(n^{4/5})]=20$ included, and then conduct a second-stage selection to fit a varying coefficient model with SCAD penalty. Table \ref{realdata} summarizes the number of genes selected by each method (Size) and mean squared prediction error (MSPE) by five-fold cross-validation. The proposed C-SIRS achieves the best performance with the sparsest model size 9 and the smallest MSPE 3.533. We show the estimated coefficient functions of the selected nine genes by C-SIRS  in Figure \ref{coef}. The identified genes is consistent with \cite{wen2018sure}, five of nine overlapped. And the estimated coefficient functions indicate the age-dependent effects of the selected genes.

\begin{table}[!h]\footnotesize
	\begin{center}
		\caption{Results for breast cancer analysis}\label{realdata}
		\setlength{\tabcolsep}{4mm}{
			\begin{tabular}{ccc}
				\hline
				Method& Size & MSPE  \\
				\hline
			    SIRS & 10 &3.694\\
				DC-SIS & 9&4.837\\
				CDC-SIS&9&3.626\\
				CC-SIS & 10&4.067 \\
			    C-SIRS &9&3.533  \\
				\hline
		\end{tabular}}
	\end{center}
\end{table}


\csection{A Brief Discussion}
In this paper, we utilize the conditional correlation between predictors and the empirical distribution function of response given exposure variables to develop a model-free sure independence screening procedure C-SIRS. It is resistant to the heavy-tailed distribution, outliers and extreme values of the response. In addition, C-SIRS is applicable for any model involving exposure variables, including generalized varying coefficient model, and any nonlinear dependent structure. It is also powerful to detect those variables with significant effects on variance components. The ranking consistency and sure screening property of C-SIRS are rendered in both theoretical properties and simulation studies. The breast cancer dataset is systematically analyzed by C-SIRS, along with the comparison with other related methods.

\scsection{Appendix: Proof of Theorems}
\scsubsection{Appendix A: Proof of Theorem \ref{theo:1}}
For easy presentation, we introduce some notations first. $\lambda_{\max}(\C)$ and $\lambda_{\min}(\C)$ denote the largest and smallest eigenvalues of a matrix $\C$, respectively. We denote $\v\v^{\trans}$ by $\v^2$ for a vector $\v$ and $\Omega_{\calA}(u,y)\defby\corr\{\x_{\calA},I(Y \le y )\mid u\}$. If we say that $a_n-b_n$ uniformly in $n$, then it means $\liminf\limits_{n\longrightarrow \infty}{a(n)-b(n)}>0$.

Without loss of generality, we assume that $E(X_k \mid u)=0$, $\var(X_k \mid u)=1$
for $k=1,\cdots,p$ and $\var\{I(Y \le y) \mid u\}=1$. Meanwhile, $\bb_{\calA}(u)$ satisfies $\bb^{\trans}_{\calA}(u)\cov(\x_{\calA}\mid u)\bb_{\calA}(u)=1$.
Then the conditional linearity condition (A1) is simplified as $E\{X_k \mid \x_{\calA}^{\trans}\bb_{\calA}(u),u\}=\cov\{X_k,\x_{\calA}^{\trans}\bb_{\calA}(u) \mid u\}\bb^{\trans}_{\calA}(u)\x_\calA$.
Following law of iterated expectations,
\beqr\nonumber
&&E\left\{X_k I(Y\le y)\mid u,y\right\}=E\left[E\left\{X_k I(Y\le y)\mid \bb^{\trans}_{\calA}(u)\x_\calA,u,y\right\}\mid u,y\right]\\
&&=E\left[E\left\{X_k \mid \bb^{\trans}_{\calA}(u)\x_\calA,u\right\}E\left\{I(Y\le y) \mid \bb^{\trans}_{\calA}(u)\x_\calA,u,y\right\} \mid u,y \right]\label{1}\\
&&=\cov\left\{X_k, \x_\calA^{\trans}\bb_\calA(u)\mid u\right\}E\left\{\bb^{\trans}_{\calA}(u)\x_\calA I(Y<y)\mid u,y\right\},\label{2}
\eeqr
where (\ref{1}) holds because of the conditional independence of $\x$ and $Y$ given $\bb^{\trans}_{\calA}(u)\x_\calA$ and $u$  in (A3) and (\ref{2}) is due to the simplified conditional linearity condition.
\beqr\nonumber
&&\max_{k \in \calA^c} \ \Omega_{k}^2(u,y)=\max_{k \in \calA^c} E^2\left\{X_k I(Y\le y)\mid u,y\right\}\\\label{3}
&&=\max_{k \in \calA^c}\left[\cov^2\left\{X_k,\x_\calA^{\trans}\bb_\calA(u)\mid u\right\}\right]\cdot E^2\left\{\bb^{\trans}_\calA(u)\x_\calA I(Y \le y) \mid u,y \right\}.
\eeqr

Then we start to deal with the first term of (\ref{3}).
\beqr\nonumber
&&\max_{k \in \calA^c}\left[\cov^2\left\{X_k,\x_\calA^{\trans}\bb_\calA(u)\mid u\right\}\right]\\\nonumber
&=&\bb^{\trans}_\calA(u)\left[\max_{k \in \calA^c}\left\{\cov\left(\x_\calA,X_k\mid u \right)\cov(X_k,\x_\calA^{\trans}\mid u)\right\}\right]\bb_\calA(u)\\\nonumber
&\le&\bb^{\trans}_\calA(u)\cov(\x_\calA,\x_{\calA^c}^{\trans} \mid u)\cov(\x_{\calA^c} ,\x_\calA^{\trans}\mid u)\bb_\calA(u)\\\nonumber
&=&\{\bb^{\trans}_\calA(u)\cov^{1/2}(\x_\calA,\x_{\calA}^{\trans}\mid u)\}\cov^{-1/2}(\x_\calA,\x_{\calA}^{\trans}\mid u)
\cov(\x_\calA,\x_{\calA^c}^{\trans} \mid u)\\\nonumber
&&\cdot \cov(\x_{\calA^c} ,\x_\calA^{\trans}\mid u)\cov^{-1/2}(\x_\calA,\x_{\calA}^{\trans}\mid u)\left\{\cov^{1/2}(\x_\calA,\x_{\calA}^{\trans}\mid u)\bb_\calA(u)\right\}\\\nonumber
&\le&\lambda_{\max}\left\{ \cov^{-1/2}(\x_\calA,\x_{\calA}^{\trans}\mid u)
\cov(\x_\calA,\x_{\calA^c}^{\trans} \mid u)\cov(\x_{\calA^c} ,\x_\calA^{\trans}\mid u)\cov^{-1/2}(\x_\calA,\x_{\calA}^{\trans}\mid u) \right\}\\ \label{4}
&\le&\lambda_{\max}\left\{
\cov(\x_\calA,\x_{\calA^c}^{\trans} \mid u)\cov(\x_{\calA^c} ,\x_\calA^{\trans}\mid u) \right\}
\cdot \lambda_{\max}\left\{\cov^{-1}(\x_\calA,\x_{\calA}^{\trans}\mid u)\right\}\\\nonumber
&=&\lambda_{\max}\left\{
\cov(\x_\calA,\x_{\calA^c}^{\trans} \mid u)\cov(\x_{\calA^c} ,\x_\calA^{\trans}\mid u) \right\}
/\lambda_{\min}\left\{\cov(\x_\calA,\x_{\calA}^{\trans}\mid u)\right\},
\eeqr
where (\ref{4}) holds because of the fact $\lambda_{\max}(\C^{\trans}\B\C)\le \lambda_{\max}(\B)\lambda_{\max}(\C^{\trans}\C)$, where the matrix $\B \ge 0$. Similarly, we can verify that
\beqrs
E^2\left\{\bb^{\trans}_\calA(u)\x_\calA I(Y \le y) \mid u,y \right\}
\le \lambda_{\max}\left[E^2\left\{\x_\calA I(Y \le y) \mid u,y \right\}\right]/\lambda_{\min}\left\{\cov(\x_\calA,\x_{\calA}^{\trans}\mid u)\right\}.
\eeqrs
These two inequations yield that
\beqrs
&&\max_{k \in \calA^c} \omega_k= \max_{k \in \calA^c} E\{\Omega_k^2(y,u)\} \le E\{\max_{k \in \calA^c} \ \Omega_k^2(y,u)\}\\
&\le&E\left[ \lambda_{\max}\left\{
\cov(\x_\calA,\x_{\calA^c}^{\trans} \mid u)\cov(\x_{\calA^c} ,\x_\calA^{\trans}\mid u) \right\} \lambda_{\max}\{\Omega_{\calA}^2(y,u)\}
/\lambda^2_{\min}\left\{\cov(\x_\calA,\x_{\calA}^{\trans}\mid u)\right\}\right],
\eeqrs
which completes the proof.\hfill$\fbox{}$

\scsubsection{Appendix B: Proof of Theorem \ref{theo:2}}
Define $E(\cdot \mid u)=g(\cdot\mid u)/f(u)$,  and
\beqrs
\wh E(X_k\mid u)=\wh g(X_k\mid u) /\wh f(u),\ \wh E\{I(Y \le y)\mid u\}=\wh g(y\mid u) /\wh f(u),
\eeqrs
where $\wh g(X_k\mid u) \defby n^{-1}\sum_{i=1}^n K_h(u_i-u)X_{ik}$,
$\wh g(y\mid u) \defby n^{-1}\sum_{i=1}^n K_h(u_i-u)I(Y_i \le y)$,
$\wh f(u) \defby n^{-1}\sum_{i=1}^n K_h(u_i-u)$ and $K_h(u) = K(u/h)/h$.
{\lemma\label{hoeffding}(Hoeffding's Inequality \citep{Wassily1963Probability}) Let $X_1,\ldots,X_n$ be independent random variables, and $\pr(a_i\le X_i\le b_i) = 1$ for $i=1,\ldots, n$. Then
	\beqrs
	\pr\left\{|\overline{X} - E(\overline{X})|\ge t\right\}
	\le 2\exp\left\{- 2n^2t^2\Big/\sum_{i=1}^n(b_i-a_i)^2\right\}, \textrm{ 	for any $t > 0$. }
	\eeqrs}
{\lemma\label{lemmaS3}\citep{Liu2014Feature} Suppose that $X$ is a random variable with $E(e^{a|X|}) < \infty$ for some $a>0$. Then  for any $M>0$, there exist positive constants $b$ and $c$ such that
	\beqrs
	\pr(|X| \ge M) \le b\exp(-cM).
	\eeqrs}
The following lemma is  a sight modified version of Theorem 3.1 of \cite{Zhu:1993} and Lemma 3.2 of \cite{zhu1996Asymptotics}.
{\lemma\label{lemma1}  Suppose that conditions (C1) to  (C5) are fulfilled,  and $\sup\limits_\x |X_k| \le  M $, then for any $\varepsilon_n >0$,
	\beqrs\nonumber
	&&\pr\left\{\sup_{u} \left|\wh g(X_k\mid u)- E \wh g(X_k\mid u)\right|\ge 8MM_Kh^{-1}n^{-1/2}\varepsilon_n\right\}
	\\\label{lemzhu-1}&&\le  3c\big(n^{1/2}/\varepsilon_n\big)^{4}\exp\left( -\varepsilon_n^2/128 \delta^2 \right)+ 4c \delta^{-8}\exp\left( -n\delta^2\right),
	\eeqrs
	where  $\delta\ge \sup\limits_{u}\left[\var\left\{K\left(\frac{\widetilde{u}-u}{h}\right)\right\}\right]^{1/2}$.}
{\lemma\label{lemma2}  Suppose that conditions (C1) to  (C5) are fulfilled,  for any $0<\gamma+\theta<1/4 $ and $0< \gamma\le m\theta$,  then we have
	\beqrs
	&&\pr\left\{\sup_{u}  \left|\wh g(X_k\mid u)-g(X_k\mid u)\right| \ge cn^{-\gamma}\right\} \le a_1 n^2\exp(-b_1n^{1/2-2\theta-2\gamma}),
	\eeqrs
	where $a_1$, $b_1$ and $c$ are some positive constants.}

\noindent{\textit{Proof of Lemma \ref{lemma2}}:}
For any $M > 0$,
\beqrs
&& \pr\left\{\sup\limits_{u}\left|\wh g(X_k\mid u)-g(X_k\mid u)\right| \ge cn^{-\gamma}\right\}\\
&&= \pr\left\{\sup\limits_{u}\left|\wh g(X_k\mid u)-g(X_k\mid u)\right| \ge cn^{-\gamma},\max |X_{ik}| \le M\right\}\\
&&+\pr\left\{\sup\limits_{u}\left|\wh g(X_k\mid u)-g(X_k\mid u)\right| \ge cn^{-\gamma},\max |X_{ik}| \ge M\right\}\\
&&\le \pr\left\{\sup\limits_{u}\left|\wh g(X_k\mid u)-g(X_k\mid u)\right| \ge cn^{-\gamma},\max |X_{ik}| \le M\right\}\\
&&+ \pr\left\{\max |X_{ik}| \ge M \ \textrm{for some} \ i\right\}=P_1(u)+P_2,
\eeqrs
where $P_1(u)$ and $P_2$ are defined in the obvious way. We derive the first term $P_1(u)$. The following arguments are all under the condition $\max |X_{ik}| \le M$, which is omitted for notation simplicity.
We   expand $g(X_k \mid u)$ in a Taylor series with the Lagrange   remainder term under Condition (C3). There exists a positive constant $c_1$ such that
\beqrs
&&\sup\limits_{u}\left|E\wh g(X_k\mid u)-g(X_k\mid u)\right|=\sup\limits_{u}\left|\int K_h\left(\widetilde{u}-u\right)\{E(X_k\mid \widetilde{u})f(\widetilde{u})-E(X_k\mid u)f(u)\}d\widetilde{u}\right|
\\&&=\sup\limits_{u}\left|\int K(t)\{E(X_k\mid u+ ht )f(u+ht )-E(X_k\mid u)f(u)\}\ dt\right|\le c_1\nu_m h^m .
\eeqrs
Since the kernel function $K$ is uniformly continuous on its compact support, we have
\beqrs
&&\sup\limits_{u}\left[\var\left\{K\left(\frac{\widetilde{u}-u}{h}\right)\right\}\right]^{1/2}
\le \sup\limits_{u} \left\{\int K^2\left(\frac{\widetilde{u}-u}{h}\right)f(\widetilde{u})d \widetilde{u}\right\}^{1/2}\le M_K
\eeqrs
Following the Lemma \ref{lemma1}, for $0< \gamma\le m\theta$,
\beqr\nonumber
&&\pr\left\{\sup_{u}  \left|\wh g(X_k\mid u)-g(X_k\mid u)\right| \ge cn^{-\gamma},\max |X_{ik}| \le M\right\}
\\\nonumber&&=\pr\left\{\sup_{u} \big|\wh g(X_k\mid u)-E \wh g(X_k\mid u)+E \wh g(X_k\mid u)-g(X_k\mid u)\big| \ge cn^{-\gamma}, \max |X_{ik}| \le M\right\}\\\nonumber
&&\le \pr\left\{\sup_{u} \big|\wh g(X_k\mid u)-E \wh g(X_k\mid u)\big|\ge cn^{-\gamma}/2,\max |X_{ik}| \le M\right\}
\\&&= O\left\{n^2\exp(-b_1n^{1-2\theta-2\gamma}/M^2)\right\},\label{1.1}
\eeqr
for some positive constant $b_1$.

Then we deal with $\pr\left\{\max |X_{ik}| \ge M \ \textrm{for some} \ i\right\} $.
Following the assumption (C5) and Lemma \ref{lemmaS3}, there exist some positive constants $t_1$ and $t_2$ such that for any $M>0$, $\pr(|X_k| \ge M) \le t_1\exp(-t_2 M)$. Then,
\beqrs
\pr\left\{\max |X_{ik}| \ge M \ \textrm{for some} \ i\right\} \le n\pr(|X_k| \ge M) \le nt_1\exp(-t_2 M) \label{1.2}
\eeqrs
Thus, together with (\ref{1.1}), we have
\beqr\nonumber
&&\pr\left\{\sup\limits_{u}\left|\wh g(X_k\mid u)-g(X_k\mid u)\right| \ge cn^{-\gamma}\right\} \le c_1n^2\exp(-b_1n^{1-2\theta-2\gamma}/M^2)+nt_1\exp(-t_2 M)\\\label{2.3}
&&\le c_1n^2\exp\{-b_1n^{1/2-2\theta-2\gamma}\cdot (n^{1/2}/M^2)\}+nt_1\exp\{-t_2 n^{1/2-2\theta-2\gamma}\cdot Mn^{2\theta+2\gamma-1/2}\}
\eeqr
Following \cite{Liu2014Feature}, we take $M=O(n^{\tau})$, where $1/2-2\theta -2\gamma< \tau<1/4$. For large $n$, we can see that
$n^{1/2}/M^2=n^{1/2-2\tau}>1, \ \textrm{and}\ Mn^{2\theta+2\gamma-1/2}=n^{\tau+2\theta+2\gamma-1/2}>1$. Now (\ref{2.3}) becomes
\beqrs
\pr\left\{\sup\limits_{u}\left|\wh g(X_k\mid u)-g(X_k\mid u)\right| \ge cn^{-\gamma}\right\} \le a_1n^2\exp(-b_1n^{1/2-2\theta-2\gamma}),
\eeqrs
where $a_1$, $b_1$ and $c$ are some positive constants. The proof is complete. \hfill$\fbox{}$
{\lemma\label{lemma5}  Suppose that conditions (C1) to  (C5) are fulfilled,  for any $0<\gamma+\theta<1/2$ and $0< \gamma\le m\theta$,  then we have
	\beqrs
	&&\pr\left\{\sup_{u}  \left|\wh f( u)-f( u)\right| \ge cn^{-\gamma}\right\}\le a_2n^2\exp(-b_2n^{1-2\theta-2\gamma}),\\
	&&\pr\left\{\sup_{u,y}  \left|\wh g(y\mid u)-g(y\mid u)\right| \ge cn^{-\gamma}\right\}\le a_3n^3\exp(-b_3n^{1-2\theta-2\gamma}),
	\eeqrs
	where $a_2$, $b_2$, $a_3$, $b_3$ and $c$ are some positive constants.}

\noindent{\textit{Proof of Lemma \ref{lemma5}}:}  We use the same technique as the proof for Lemma \ref{lemma2}. According to Theorem 3.1 of \cite{Zhu:1993} and Lemma 3.1 of \cite{zhu1996Asymptotics}, for any $\varepsilon_n >0$,	
\beqr\nonumber
&&\pr\left\{\sup_{u} \left|\wh f(u)- E \wh f( u)\right|\ge 8M_Kh^{-1}n^{-1/2}\varepsilon_n\right\}
\\\label{lemzhu-3}&&\le  3A\big(n^{1/2}/\varepsilon_n\big)^{4}\exp\left( -\varepsilon_n^2/128 \delta_{1}^2 \right)+ 4A \delta_{1}^{-8}\exp\left( -n\delta_{1}^2\right),
\\\nonumber&&\pr\left\{\sup_{u,y} \left|\wh g(y\mid u)- E \wh g(y\mid u)\right|\ge 8M_Kh^{-1}n^{-1/2}\varepsilon_n\right\}
\\\label{lemzhu-2}&&\le  3A\big(n^{1/2}/\varepsilon_n\big)^{6}\exp\left( -\varepsilon_n^2/128 \delta_{2}^2 \right)+ 4A \delta_{2}^{-12}\exp\left( -n\delta_{2}^2\right),
\eeqr
where  $\delta_1\ge \sup\limits_{u}\left[\var\left\{K\left(\frac{\widetilde{u}-u}{h}\right)\right\}\right]^{1/2}$ and $\delta_2\ge \sup\limits_{u,y} \left[\var\left\{K\left(\frac{\widetilde{u}-u}{h}\right)I(Y \le y)\right\}\right]^{1/2}$.  And $A$ is  a generic constant and  may take different values at different places.
We   expand $f(u)$ and $g(y \mid u)$  in a Taylor series with the Lagrange   remainder term under Condition (C3). There exists some positive constants $c_2$ and $c_3$ such that
\beqrs
&&\sup\limits_{u}\left|E \wh f(u)-f(u)\right|=\sup\limits_{u}\left|\int K_h\left(\widetilde{u}-u\right)\left\{f(\widetilde{u})-f(u)\right\}d\widetilde{u}\right|
\\&&=\sup\limits_{u}\left|\int K\left(t\right)\{f(u+ht)-f(u)\}\ dt\right|\le c_2\nu_m h^m,\\
&&\sup\limits_{u,y}\left|E\wh g(y\mid u)-g(y\mid u)\right|=\sup\limits_{u}\left|\int K_h\left(\widetilde{u}-u\right)\{F(y\mid \widetilde{u})f(\widetilde{u})-F(y\mid u)f(u)\}d\widetilde{u}\right|
\\&&=\sup\limits_{u,y}\left|\int K(t)\{F(y\mid u+ ht )f(u+ht )-F(y\mid u)f(u)\}\ dt\right|\le c_3\nu_m h^m .
\eeqrs
Since the kernel function $K$ is uniformly continuous on its compact support, we have
\beqrs
&&\sup\limits_{u,y}\left[\var\left\{K\left(\frac{\widetilde{u}-u}{h}\right)I(Y \le y)\right\}\right]^{1/2}
\le \sup\limits_{u}\left\{\int K^2\left(\frac{\widetilde{u}-u}{h}\right)f(\widetilde{u})d \widetilde{u}\right\}^{1/2}\le M_K
\eeqrs
Thus for $0< \gamma\le m\theta$,
\beqr\nonumber
&&\pr\left\{\sup_{u}  \left|\wh f(u)-f(u)\right| \ge cn^{-\gamma}\right\}=\pr\left\{\sup_{u} \left|\wh f(u)-E \wh f(u)+E \wh f(u)-f(u)\right| \ge cn^{-\gamma}\right\}
\\&&\le \pr\left\{\sup_{u} \left|\wh f(u)-E \wh f(u) \right|\ge cn^{-\gamma}/2\right\}=O\left\{n^2\exp(-b_2n^{1-2\theta-2\gamma})\right\},
\\\nonumber&&\pr\left\{\sup_{u,y}  \left|\wh g(y\mid u)-g(y\mid u)\right| \ge cn^{-\gamma}\right\}
=\pr\left\{\sup_{u,y} \big|\wh g(y\mid u)-E \wh g(y\mid u)+E \wh g(y\mid u)\right.\\\nonumber
&&\left.-g(y\mid u)\big| \ge cn^{-\gamma}\right\}
\le \pr\left\{\sup_{u,y} \big|\wh g(y\mid u)-E \wh g(y\mid u)\big|\ge cn^{-\gamma}/2 \right\}
\\&&= O\left\{n^3\exp(-b_3n^{1-2\theta-2\gamma})\right\},\label{2.1}
\eeqr
where $b_2$, $b_3$ and $c$ are some positive constants. The proof is complete. \hfill$\fbox{}$

{\lemma\label{lemma6} \citep{Liu2014Feature} Suppose $T(u,y)$ and $S(u,y)$ are two uniformly-bounded functions of $u$ and $y$. For any given $u$ and $y$, $\wh T(u,y)$ and $\wh S(u,y) $ are estimates of $T(u,y)$ and $S(u,y)$ based on a sample with size $n$. For any $0<\gamma+\theta<1/4$ and $0< \gamma\le m\theta$, suppose that
	\beqrs
	&&\pr\left\{\sup_{u,y}  \left|\wh  T(u,y)-T(u,y)\right| \ge cn^{-\gamma}\right\}\le a_4n^3\exp(-b_4n^{1/2-2\theta-2\gamma}),\\
	&&\pr\left\{\sup_{u,y}  \left|\wh   S(u,y)- S(u,y)\right| \ge cn^{-\gamma}\right\}\le a_5n^3\exp(-b_5n^{1/2-2\theta-2\gamma}).
	\eeqrs
	where $a_4$, $b_4$, $a_5$, $b_5$ and $c$ are some positive constants.Then we have
	\beqrs
	&&\pr\left\{\sup_{u,y}  \left|\wh  T(u,y) \wh S(u,y)-T(u,y) S(u,y)\right| \ge cn^{-\gamma}\right\}\le a_6n^3\exp(-b_6n^{1/2-2\theta-2\gamma}),\\
	&&\pr\left\{\sup_{u,y}  \left|\wh  T(u,y) /\wh S(u,y)-T(u,y) /S(u,y)\right| \ge cn^{-\gamma}\right\}\le a_7n^3\exp(-b_7n^{1/2-2\theta-2\gamma}),\\
	&&\pr\left\{\sup_{u,y}  \left|\{\wh  T(u,y) \wh S(u,y)\}-\{T(u,y)- S(u,y)\}\right| \ge cn^{-\gamma}\right\}\le a_8n^3\exp(-b_8n^{1/2-2\theta-2\gamma}),
	\eeqrs
	where $a_i$, $b_i$ for $i=6,7,8$ and $c$ are some positive constants.}

\noindent{\textit{Proof of Theorem  \ref{theo:2}}:} We divide the proof into two steps.
\\\noindent {\bf Step 1}. We first prove that, under conditions (C1)-(C5), for any $0< \gamma + \theta  \le 1/4$ and $0< \gamma\le m\theta$, there exists some positive constant $c$ such that
\beqrs
\pr\left(|\wh \omega_k-  \omega_k|\ge c n^{-\gamma}\right)\le cn^3\exp(-cn^{1/2-2\gamma-2\theta}).
\eeqrs

We define $\Delta_1$ and $\Delta_2$ as follows,
\beqrs
\Delta_1&=&n^{-2}\sum_{i=1}^n\sum_{j=1}^n \left[\wh \corr^2\{X_k,I(Y\le Y_j)\mid u_i\}-\corr^2\{X_k,I(Y\le Y_j)\mid u_i\}\right];\\
\Delta_2&=&n^{-2}\sum_{i=1}^n\sum_{j=1}^n \left[\corr^2\{X_k,I(Y\le Y_j)\mid u_i\}-\omega_k\right].
\eeqrs
Then we have $\pr\left(|\wh \omega_k-  \omega_k|\ge c n^{-\gamma}\right)\le \pr\left(|\Delta_1|\ge c n^{-\gamma}/2\right)+ \pr\left(|\Delta_2|\ge c n^{-\gamma}/2\right)$.
We deal with the first part of the summation.
\beqr\nonumber
&&\pr\left(|\Delta_1|\ge c n^{-\gamma}/2\right)
\le\pr\Big(n^{-2}\sum_{i=1}^n\sum_{j=1}^n \left|\wh \corr^2\{X_k,I(Y\le Y_j)\mid u_i\} \right.-
\\\nonumber&&\left.\corr^2\{X_k,I(Y\le Y_j)\mid u_i\}\right|\ge cn^{-\gamma}/2\Big)
\le\pr\Big(\sup_{u,y} \left|\wh \corr^2\{X_k,I(Y\le y)\mid u\} \right.-
\\&&\left.\corr^2\{X_k,I(Y\le y)\mid u\}\right|\ge cn^{-\gamma}/2\Big).\label{2.8}
\eeqr
For notation clarity, we define $\wh g(X_k,y\mid u) \defby n^{-1}\sum_{i=1}^n K_h(u_i-u)X_{ik}I(Y_i \le y)$  and $\wh g(X^2_k\mid u)\defby n^{-1}\sum_{i=1}^n K_h(u_i-u)X^2_{ik}$. Then $\corr^2\{X_k,I(Y\le y)\mid u\}$ can be written as
\beqrs
\wh \corr^2\{X_k,I(Y\le y)\mid u\} =\frac{\left\{\wh g(X_k,y\mid u)\wh f(u)-\wh g(X_k\mid u)\wh g(y\mid u)\right\}^2}{\left\{\wh g(X_k^2\mid u)\wh f(u)-\wh g^2(X_k\mid u)\right\} \left\{\wh g(y\mid u)\wh f(u)-\wh g^2(y\mid u)\right\}}
\eeqrs
Similar to the proof of Lemma \ref{lemma2}, we can obtain
\beqrs
&&\pr\left\{\sup_{u}  \left|\wh g(X_k\mid u)-g(X_k\mid u)\right| \ge cn^{-\gamma}\right\} \le a_9 n^2\exp(-b_9n^{1/2-2\gamma-2\theta}),
\eeqrs
Together with Lemmas \ref{lemma2}, \ref{lemma5}  and \ref{lemma6},  it is clear that
\beqrs
\pr\Big(\sup\limits_{u,y} \left|\wh \corr^2\{X_k,I(Y\le y)\mid u\} -\corr^2\{X_k,I(Y\le y)\mid u\}\right|\ge cn^{-\gamma}\Big) \le cn^3\exp(-cn^{1/2-2\gamma-2\theta}).
\eeqrs
Thus (\ref{2.8}) becomes $\pr\left(|\Delta_1|\ge c n^{-\gamma}/2\right) \le cn^3\exp(-cn^{1/2-2\gamma-2\theta})$. With Hoeffding's inequality in Lemma \ref{hoeffding}, we can show that $\pr \left( | \Delta_2 | \ge cn^{-\gamma}/2\right) \le 2 \exp(-c^2n^{1-2\gamma}/2)$. Thus, there exists some positive constants $c$ such that
\beqrs
\pr\left(|\wh \omega_k-  \omega_k|\ge c n^{-\gamma}\right)\le cn^3\exp(-cn^{1/2-2\gamma-2\theta}).
\eeqrs


\noindent {\bf Step 2}. Assume the condition $\min\limits_{k\in \mathcal{A}} \ \omega_k \geq 2cn^{-\gamma}$. We prove that
\beqrs
\pr(\cal{A} \subseteq \wh{\cal{A}})  &\ge& 1-O\left\{n^{3}|\mathcal{A}|  \exp( -cn^{1/2-2\gamma-2\theta })\right\}.
\eeqrs
If $\mathcal{A} \nsubseteq \widehat{\mathcal{A}}$, there must exist some $j\in\mathcal{A}$ such that $\wh \omega_j<cn^{-\gamma}$. Under the condition $\min\limits_{k\in \mathcal{A}} \ \omega_k \geq 2cn^{-\gamma}$,   we have $\left|\wh \omega_j-\omega_j\right|\geq cn^{-\gamma}$ for this particular $j$, which implies
\beqrs
\{\mathcal{A}\nsubseteq \wh{\cal{A}}  \} \subseteq \{\left|\wh \omega_j - \omega_j\right| \geq cn^{-\gamma},\,\text{for some } j\in \mathcal{A}\}.
\eeqrs
Then, it is clear that
\beqrs
\pr(\cal{A} \subseteq \wh{\cal{A}})  &\ge& 1-\pr\left\{\left|\wh \omega_j - \omega_j\right| \geq cn^{-\gamma},\,\text{for some } j\in \mathcal{A}\right\}
\\&\ge& 1-|\mathcal{A}|\max_{j\in \mathcal{A}}\pr\{ |\wh \omega_j - \omega_j|\ge cn^{-\gamma}\}
\\ &\ge& 1- O\left\{n^{3}|\mathcal{A}|  \exp( -cn^{1/2-2\gamma-2\theta })\right\},
\eeqrs
which implies the desired conclusion. \hfill$\fbox{}$

\scsubsection{Appendix C: Proof of Theorem \ref{theo:3}}
The conditions (A1)-(A3) illustrate that $\min\limits_{k \in \cal{A}} \ \omega_k -\max\limits_{k \in \calA^c} \ \omega_k \ge 0$.Thus, there exists some $\delta>0$ such that $\min\limits_{k \in \cal{A}} \ \omega_k -\max\limits_{k \in \calA^c}\ \omega_k=\delta$. Then we have
\beqrs
&&\pr\left\{\min\limits_{k \in \cal{A}}  \wh\omega_k \le\max\limits_{k \in \calA^c} \wh\omega_k\right\}
\le \pr \left\{\min\limits_{k \in \cal{A}} \wh\omega_k -\min\limits_{k \in \cal{A}} \omega_k+ \delta
\le \max\limits_{k \in \calA^c} \wh \omega_k -\max\limits_{k \in \calA^c}  \omega_k\right\}
\\&&\le \pr \left\{\left|(\min\limits_{k \in \cal{A}} \wh \omega_k-\max\limits_{k \in \calA^c} \wh \omega_k)-(\min\limits_{k \in \cal{A}}  \omega_k -\max\limits_{k \in \calA^c}  \omega_k)\right| \ge \delta\right\}
\\&&\le \pr \left\{2\max\limits_{1 \le k \le p} \left|\wh \omega_k - \omega_k\right| \ge \delta\right\}\le O\left\{p\,n^{3} \exp\left( -cn^{1/2-2\theta }\right)\right\},
\eeqrs
Then by Fatou's Lemma,
\beqrs
\pr\left\{\liminf\limits_{n\longrightarrow \infty} \left(\min\limits_{k \in \cal{A}} \wh\omega_k -\max\limits_{k \in \calA^c} \wh\omega_k\right) \le 0 \right\} \le \lim\limits_{n\rightarrow \infty} \pr\left(\min\limits_{k \in \cal{A}}  \wh \omega_k -\max\limits_{k \in \calA^c} \wh \omega_k\le 0\right)=0
\eeqrs
In other words, we have
$\pr\left\{\liminf\limits_{n\longrightarrow \infty} \left(\min\limits_{k \in \cal{A}} \wh\omega_k -\max\limits_{k \in \calA^c} \wh\omega_k\right) > 0 \right\}=1. $\hfill$\fbox{}$

\begin{singlespace}
\scsection{References}

\begin{description}
\newcommand{\enquote}[1]{``#1''}
\expandafter\ifx\csname
natexlab\endcsname\relax\def\natexlab#1{#1}\fi


%



\bibitem[Chin et~al., 2006]{chin2006genomic}
Chin, K., DeVries, S., Fridlyand, J., Spellman, P.~T., Roydasgupta, R., Kuo,
  W.-L., Lapuk, A., Neve, R.~M., Qian, Z., Ryder, T., et~al. (2006).
\newblock Genomic and transcriptional aberrations linked to breast cancer
  pathophysiologies.
\newblock {\em Cancer Cell}, 10(6):529--541.

\bibitem[DeSantis et~al., 2017]{desantis2017breast}
DeSantis, C.~E., Ma, J., Goding~Sauer, A., Newman, L.~A., and Jemal, A. (2017).
\newblock Breast cancer statistics, 2017, racial disparity in mortality by
  state.
\newblock {\em CA: A Cancer Journal for Clinicians}, 67(6):439--448.


\bibitem[Fan et~al., 2014]{Fan2014Nonparametric}
Fan, J., Ma, Y., and Dai, W. (2014).
\newblock Nonparametric independence screening in sparse ultra-high dimensional
  varying coefficient models.
\newblock {\em Journal of the American Statistical Association},
  109(507):1270--1284.

\bibitem[Fan et~al., 2011]{FanFengSong:2011}
Fan, J., Feng, Y., and Song, R. (2011).
\newblock Nonparametric independence screening in sparse ultra-high dimensional
  additive models.
\newblock {\em Journal of the American Statistical Association},
  106(494):544--557.


\bibitem[Fan and Lv, 2008]{fan2008sure}
Fan, J. and Lv, J. (2008).
\newblock Sure independence screening for ultrahigh dimensional feature space.
\newblock {\em Journal of the Royal Statistical Society: Series B},
  70(5):849--911.


\bibitem[Fan and Song, 2010]{fan2010sure}
Fan, J. and Song, R. (2010).
\newblock Sure independence screening in generalized linear models with
  \protect{NP}-dimensionality.
\newblock {\em The Annals of Statistics}, 38(6):3567--3604.


\bibitem[He et~al., 2013]{he2013quantile}
He, X., Wang, L., Hong, H.~G. (2013).
\newblock Quantile-adaptive model-free variable screening for high-dimensional
 heterogeneous data.
\newblock {\em The Annals of Statistics}, 41(1):342--369.

\bibitem[Hoeffding, 1963]{Wassily1963Probability}
Hoeffding, W. (1963).
\newblock Probability inequalities for sums of bounded random variables.
\newblock {\em Journal of the American Statistical Association},
  58(301):13--30.

\bibitem[Li et~al., 2012a]{li2012robust}
Li, G., Peng, H., Zhang, J., and Zhu, L. (2012a).
\newblock Robust rank correlation based screening.
\newblock {\em The Annals of Statistics}, 40(3):1846--1877.

\bibitem[Li et~al., 2012b]{li2012feature}
Li, R., Zhong, W., and Zhu, L. (2012b).
\newblock Feature screening via distance correlation learning.
\newblock {\em Journal of the American Statistical Association},
  107(499):1129--1139.

\bibitem[Liu et~al., 2014]{Liu2014Feature}
Liu, J., Li, R., and Wu, R. (2014).
\newblock Feature selection for varying coefficient models with ultrahigh
  dimensional covariates.
\newblock {\em Journal of the American Statistical Association},
  109(505):266--274.

 \bibitem[Liu et~al., 2015]{Liu2015Aselective}
 Liu, J., Zhong, W. and Li, R. (2015).
  \newblock A selective overview of feature screening for ultrahigh-dimensional data.
\newblock{\em Science China Mathematics}, 58(10):2033--2054.



 \bibitem[McPherson et~al., 2000]{mcpherson2000abc}
McPherson, K., Steel, C., and Dixon, J. (2000).
\newblock Abc of breast diseases: breast cancer¡ªepidemiology, risk factors,
  and genetics.
\newblock {\em BMJ: British Medical Journal}, 321(7261):624-628.

\bibitem[Wang et~al., 2015]{wang2015conditional}
Wang, X., Pan, W., Hu, W., Tian, Y., and Zhang, H. (2015).
\newblock Conditional distance correlation.
\newblock {\em Journal of the American Statistical Association},
  110(512):1726--1734.

  \bibitem[Wen et~al., 2018]{wen2018sure}
  Wen, C., Pan, W., Huang, M., and Wang, X. (2018).
  \newblock Sure independence screening adjusted for confounding covariates with
  ultrahigh dimensional data.
  \newblock {\em Statistica Sinica}, 28:293--317.

  \bibitem[Witten et~al., 2009]{witten2009penalized}
Witten, D.~M., Tibshirani, R., and Hastie, T. (2009).
\newblock A penalized matrix decomposition, with applications to sparse
  principal components and canonical correlation analysis.
\newblock {\em Biostatistics}, 10(3):515--534.

   \bibitem[\protect\astroncite{PMA}]{PMA:2013}
  Witten, D.~M., Tibshirani, R., Gross, S. and Narasimhan, B.
  \newblock PMA: Penalized multivariate analysis. R package version 1.0.9, 2013.
  \newblock Available from: http://CRAN.R-project.org/package=PMA.


\bibitem[Zhu et~al., 2011]{zhu2011model}
Zhu, L.~P., Li, L., Li, R., and Zhu, L.~X. (2011).
\newblock Model-free feature screening for ultrahigh-dimensional data.
\newblock {\em Journal of the American Statistical Association},
  106(496):1464--1475.
\bibitem[{Zhu(1993)}]{Zhu:1993}
Zhu, L. X. (1993).
\newblock Convergence rates of the empirical processes indexed by the classes of
functions with applications.
\newblock {\em Journal of Systems Science and Mathematical Sciences},
  13(1):33--41.
  \bibitem[Zhu and Fang, 1996]{zhu1996Asymptotics}
  Zhu, L. X. and Fang, K. T. (2007).
  \newblock Asymptotics for kernel estimate of sliced inverse regression.
  \newblock {\em The Annals of Statistics}, 24(3):1053--1068.
\end{description}
\end{singlespace}
\end{document}